\documentclass[aps,prl,twocolumn]{revtex4-1}
\usepackage{graphicx}
\usepackage{bm}
\usepackage{amssymb}
\begin{document}
\title{Conformal Invariance in Inverse Turbulent Cascades}
\author{G. Falkovich$^1$ and S. Musacchio$^2$}
\affiliation{
$^1$ Weizmann Institute of Science, Rehovot,Israel\\
$^2$ SCNRS, Lab. J.A. Dieudonn\'e UMR 6621,Parc Valrose, 06108 Nice (France)}
\date{\today}
\begin{abstract}
We study statistical properties of turbulent inverse cascades in a class of
nonlinear models describing a scalar field transported by a two-dimensional
incompressible flow. The class is characterized by a linear relation between
the transported field and the velocity, and include several cases of physical
interest, such as Navier-Stokes, surface quasi-geostrophic and
Charney-Hasegawa-Mima equations. We find that some statistical properties of
the inverse turbulent cascades in such systems are conformal invariant. In
particular, the zero-isolines of the scalar field are statistically equivalent
to conformal invariant curves within the resolution of our numerics. We show
that the choice of the conformal class is determined by the properties of a
transporting velocity rather than those of a transported field and discover a
phase transition when the velocity turns from a large-scale field to a
small-scale one.
\end{abstract}
\maketitle
%%%%%%%%%%%%%%%%%%%%%%%%%%%%%%%%%%%%%%%%%%%%%%%%%%%%%%%%%%%%%%%%%
%\section{Introduction}
Exceptional role of conformal invariance in theoretical physics
stems from the fact that most of non-trivial exact solutions of
dynamical and statistical models can be traced to the existence of
this symmetry. One of the most remarkable recent advances in
mathematics was the discovery of Schramm-Loewner Evolution (SLE) and
of the bridges it builds between different branches of physics
\cite{Schramm,C05,BB06}. SLE is a class of fractal random curves
that can be mapped into a one-dimensional Brownian walk and thus
have conformal invariant statistics. SLE curves appear at
two-dimensional (2d) critical phenomena as cluster boundaries, thus
revealing a statistical geometry of conformal field theories.

In equilibrium, the statistical weight of a state does not depend on how the
state was created. If a system is driven away from equilibrium by an external
force then the probability of a given configuration depends generally on the
history and on the statistics of the driving force. Turbulence statistics are
thus generally force-dependent. All the more surprising was then the
experimental discovery that the isolines of vorticity in the 2d Navier-Stokes
turbulence and of temperature in the Surface Quasi-Geostrophic (SQG) turbulence
belong to SLE \cite{BBCF06,BBCF07}. That means that at least a part of
turbulence statistics could be described in terms of a conformal field theory
like equilibrium critical phenomena.  In particular, nodal vorticity lines
happen to be equivalent to the boundaries of percolation clusters
\cite{BBCF06}, while the iso-temperature lines in SQG are equivalent to the
domain walls of SO(2) model (that of a 2d gaussian free field) \cite{BBCF07}.
Having only two examples leaves wide open possibilities for different
interpretations and hypothesis, particularly trying to relate the scaling and
properties of the bulk field to the choice of the curve class for its isolines
\cite{Fal}. Here we study additional models from the family and show that the
SLE class is actually sensitive to the type of dynamics (i.e. velocity) rather
than to the type of a field that is carried; that sensitivity manifested
dramatically by the phase transition we discover.

The class of models we investigate has been introduced in \cite{Const,PHS94}.
It describes the evolution of a scalar field $\theta$ transported by an
incompressible two-dimensional velocity ${\bm u} = (\partial_y \psi,
-\partial_x \psi)$, expressed via the stream function $\psi$. The scalar field
$\theta$ is ``active'' because it is linearly related to $\psi$ and ${\bm u}$.
In Fourier space the relation reads: $\theta({\bm k}) = |{\bm k}|^{m} \psi({\bm
k})$. The system is thus governed by the equation
\begin{equation}
\partial_t \theta + ({\bm u}\nabla) \theta=\partial_t \theta +
\{\theta,\psi\} = F + D \;,
%\;\;\theta({\bm k}) = |{\bm k}|^{\alpha} \psi({\bm k})
\label{eq:1}
\end{equation}
where $\{\theta,\psi\}=\theta_x\psi_y-\theta_y\psi_x$, $F$ and $D$
are external forcing and dissipation respectively.
%The parameter $m$ controls the statistical properties of the scalar field.
Different values of $m$ give different well-known hydrodynamic
equations. For $m = 2$ one obtains two-dimensional Navier-Stokes
(NS) equation, $\theta$ being the vorticity. For $m = 1 $ the
field $\theta$ represents the temperature in SQG
 turbulence. Finally, for $m = -2$
the model corresponds to that derived by  Charney and Oboukhov for
waves in rotating fluids and by Hasegawa and Mima for drift waves
in magnetized plasma in the limit of vanishing Rossby radius (ion
Larmor radius for plasma physics).

At all values of $m$ equation~(\ref{eq:1}) possesses two positive-definite
invariants for $F=D=0$, namely $Z=\int \theta^2 d{\bm x} $ and $E = \int \theta
\psi d{\bm x}/2$. When the system is forced by an external source of scalar
fluctuations $F$, with a correlation length $\ell_f \sim 1/k_f$, the existence
of two conserved quantities causes double turbulent cascade. The sign of $m$
determines the direction of the cascades. For $m >0$ the ``energy'' $E$ is
transferred toward large scales $\ell > \ell_f$ giving rise to an inverse
cascade, and the ``enstrophy'' $Z$ flows toward small scales. The cascades are
reversed for  $m <0$.

%%%%%%%%%%%%%%%%%%%%%%%%%%%%%%%%%%%%%%%%%%%%%%%%%%%%%%%%%%%
\begin{figure} [ht!]
\centerline{
\includegraphics[scale=0.7,draft=false]{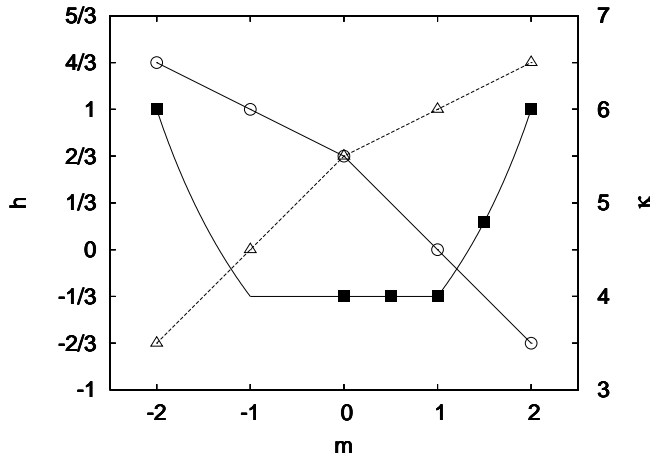}}
\caption{Scaling exponents $h$ of the scalar field $\theta$
(circles) and stream function $\phi$ (triangles), and universality
class $\kappa$ (squares) for various values of $m$.} \label{fig:1}
\end{figure}
%%%%%%%%%%%%%%%%%%%%%%%%%%%%%%%%%%%%%%%%%%%%%%%%%%%%%%%%%%%

Here we focus on the  range of scales corresponding to an inverse cascade.
Dimensional argument  based on the assumption of scale-independence of the flux
of energy in the inverse cascade (for $m > 0$) gives  the scaling exponent
$h=(2-2m)/3$ for the the increments $\delta_r \theta = \theta({\bm x}+{\bm r})
-\theta({\bm x}) \sim r^h$ and $h = (2+m)/3$ for the stream function $\psi$.
For $m < 0$ similar argument gives the scaling exponent $h=(2-m)/3$ for the
field $\theta$, and $h = (2+2m)/3$ for the stream function $\psi$. Our direct
numerical simulations support these predictions. The scaling exponents of the
fields $\theta$ and $\psi$ are shown in Figure~\ref{fig:1} as a function of
$m$. When $m\to-m$, fields $\theta$ and $\psi$ exchange their scaling exponents
(and the respective cascades change directions).

The first remarkable discovery of conformal invariance in turbulence
has been made for the zero-vorticity lines in Navier-Stokes
turbulence i.e. for $m=2$ \cite{BBCF06}. Zero-vorticity regions
correspond to $\Delta\psi=0$. i.e. to  a harmonic stream-function
and are invariant with respect to conformal transformations (which
thus map streamlines into themselves). One may  think that conformal
invariance of zero-vorticity lines is a remnant of the invariance of
zero-vorticity domains and is peculiar  for $m=2$. However,
conformal invariance of the isolines was then discovered for $m=1$
\cite{BBCF06}, where one does not recognize an analogous property of
zero-$\theta$ domains. It is then tempting to relate conformal
invariance to the properties of $\theta$ which are common for all
$m$. The main property seems to be the fact that $\theta$ is a
Lagrangian invariant of the flow and determines the symplectic
structure. For example, if we denote ${\bf R}=(X,Y)$ the initial
(Lagrangian) coordinates of the fluid particles then the extremum of
the action $I=\int S(t)\,dt$ with
\begin{eqnarray}&&S_2({\bf r})=\int \theta({\bf R})x({\bf
R},t)\dot y({\bf R},t)\,d{\bf R}\nonumber\\&&-{1\over2}\int\theta({\bf
R}_1)\theta({\bf R}_2)\ln|{\bf r}({\bf R}_1)-{\bf r}({\bf R}_2)|\, d{\bf
R}_1d{\bf R}_2 \label{action1}\end{eqnarray} gives $\dot x=\partial_y \psi$ and
$\dot y=-\partial_x\psi$ which is equivalent to (\ref{eq:1}) for $m=2$.
Generally,
\begin{eqnarray}&&S_m({\bf r})=\int \theta({\bf R})x({\bf
R},t)\dot y({\bf R},t)\,d{\bf
R}\nonumber\\&&-{1\over2}\int\theta({\bf R}_1)\theta({\bf
R}_2)|{\bf r}({\bf R}_1)-{\bf r}({\bf R}_2)|^{m-2}\, d{\bf
R}_1d{\bf R}_2\ .\label{action2}\end{eqnarray}
%For steady flows, one can define canonical structure independent of $\theta$
% with $\psi(x,y)$ being the Hamiltonian.
In other words, the energy $E= \int\psi({\bf r})\theta({\bf
r})\,d{\bf r}/2=\int\theta({\bf r}_1)\theta({\bf r}_2)|{\bf r}({\bf
R}_1)-{\bf r}({\bf R}_2)|^{m-2}\, d{\bf r}_1d{\bf r}_2/2$ is the
Hamiltonian.  It is tempting to conjecture \cite{Gaw} that
zero-$\theta$ lines are special since the Hamiltonian description is
singular (non-invertible) there. However, at negative $m$, $\theta$
is a large-scale field and its isolines are not fractal (have
dimensionality 1). It is then natural to study the properties of the
isolines of $\psi$ which are fractal now. We show below that  at
$m=-2$ the isolines of $\psi$ seem to have the same statistical
properties as the isolines of $\theta$ at $m=2$, despite the fact
that $\psi$ is not a Lagrangian invariant and the equation has no
symmetry $m\to-m$.

%%%%%%%%%%%%%%%%%%%%%%%%%%%%%%%%%%%%%%%%%%%%%%%%%%%%%%%%%%%
\begin{figure} [t!]
\centerline{
\includegraphics[scale=0.7,draft=false]{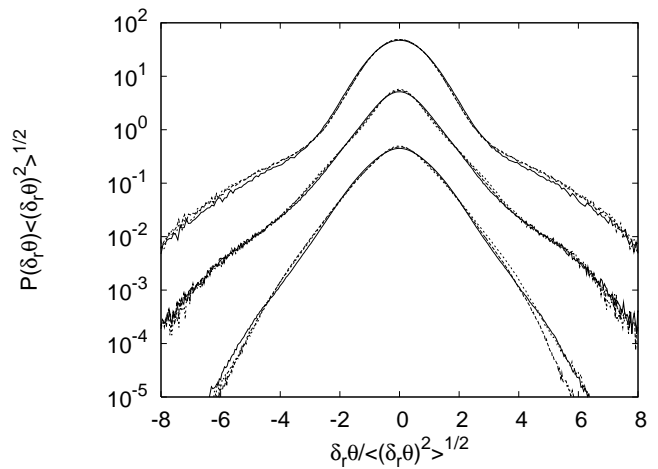}}
\caption{Pdfs of scalar increments $\delta \theta(r)$ at different
distances $r$ within the inertial range for $m=3/2$, $m=1/2$,
$m\to0$ (from top to bottom)} \label{fig:2}
\end{figure}
%%%%%%%%%%%%%%%%%%%%%%%%%%%%%%%%%%%%%%%%%%%%%%%%%%%%%%%%%%%

%In the following we will focus on the case $m > 0$.
To investigate the statistical properties of the scalar field
$\theta$ we solved numerically eq.~\ref{eq:1} on a doubly periodic
square domain of size $L=2\pi$ at different resolution $N^2 =
1024^2, 2048^2$. The scalar fluctuations are sustained by a
Gaussian, $\delta$-correlated in time, random forcing, peaked
around wavenumber $k_f = 100$. Dealiasing cutoff is set to $k_t=
N/3$. Time evolution was computed by means of a second-order
Runge--Kutta scheme, with implicit handling of the linear
dissipative terms. The direct cascade of enstrophy is halted at
wavenumbers $k>k_f$ by means of a hyper-viscous damping
$(-1)^{p-1}\nu_{p-1}\nabla^{2\,p}\,\theta$ of order $p=8$.
Statistically steady state in the inverse cascade is obtained by
removing the energy at large scales with a linear friction term
$-\eta \theta$. Note that for $m > 0$ the characteristic times of
the inverse cascade process scales as $\tau_{\ell} \sim
\ell^{(4-m)/3}$ i.e. the cascade slows down as $m$ goes to zero.
This phenomenon limits the resolution achievable in numerical
simulations.

%%%%%%%%%%%%%%%%%%%%%%%%%%%%%%%%%%%%%%%%%%%%%%%%%%%%%%%%%
\begin{figure} [th!]
\centerline{
\includegraphics[scale=0.7,draft=false]{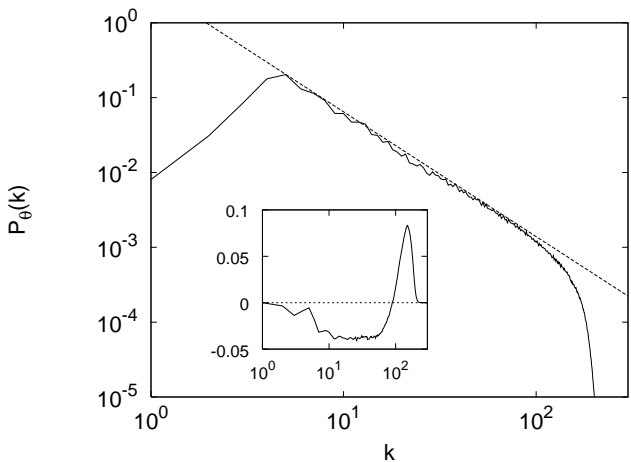}}
\caption{Power spectrum of the scalar field $\theta$ for $m=1/2$
Dashed line is the slope $k^{-5/3}$. The inset shows the energy
flux.} \label{fig:3}
\end{figure}
%%%%%%%%%%%%%%%%%%%%%%%%%%%%%%%%%%%%%%%%%%%%%%%%%%%%%%%%%

The scalar field resulting from numerical simulations with $0<m
\le 2$ is scale invariant, as confirmed by the perfect collapse of
the probability distribution functions (pdfs) of scalar increments
$\delta_r \theta$ for different $r$, see e.g. Figure~\ref{fig:2}.
Note that the pdfs are non-Gaussian. The scalar field also has a
power-law spectrum for $k<k_f$ in agreement with the prediction:
\begin{equation}
P_\theta(k) = C_{m} \epsilon^{2/3} k^\zeta \label{eq:2}
\end{equation}
where $\zeta = (4m -7)/3$, and $\epsilon$ is the flux of energy
(see e.g. Figure~\ref{fig:3}).

%\section{Asymptotic model $m \to 0$}

\begin{figure} [th!]
\centerline{
\includegraphics[scale=0.7,draft=false]{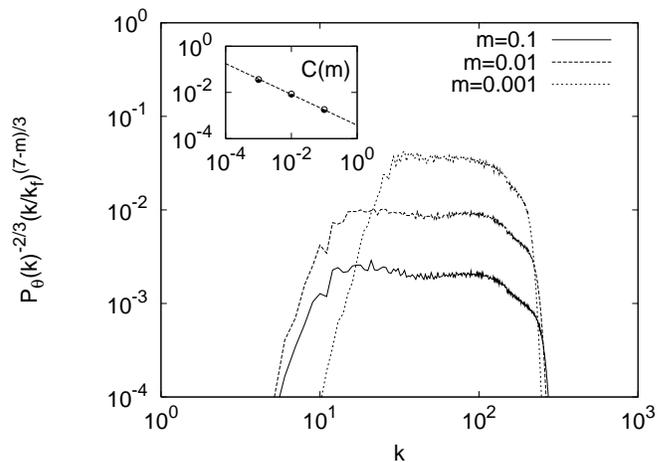}}
\caption{Compensated power spectrum of the scalar field $\theta$
for $m \to 0$. Here $m = 10^{-1},10^{-2},10^{-3}$. In the inset it
is shown the power law behavior of the amplitude $C(m) \sim
m^{-2/3}$.} \label{fig:4}
\end{figure}

The limit $m \to 0$ of the active scalar model is singular. Indeed
for $m=0$ the two fields $\theta$ and $\psi$ coincide, and the
advection term $\{\theta,\psi\}$ in eq.~\ref{eq:1} vanishes.
Therefore no turbulent state can be produced and the field
$\theta$ is simply determined by local balance between forcing and
the dissipation at exactly  $m=0$. Conversely, for arbitrary small
values of $m$ we find a turbulent cascade with power law spectrum
in agreement with eq.~\ref{eq:2} (see Figure~\ref{fig:4}). As the
parameter $m$ goes to zero, the amplitude of the scalar field
diverges, to compensate for the less efficient transfer of energy
in the cascade. This is signalled by the power law behavior of the
analogous of Kolmogorov's constant for the spectrum $C(m) \sim
m^{-2/3}$ (see inset of Figure~\ref{fig:4}).

%%%%%%%%%%%%%%%%%%%%%%%%%%%%%%%%%%%%%%%%%%%%%%%%%%%%%%
%%%%%%%%%%%%%%%%%%%%%%%%%%%%%%%%%%%%%%%%%%%%%%%%%%%%%%%%%

\begin{figure} [th!]
\centerline{
\includegraphics[scale=0.7,draft=false]{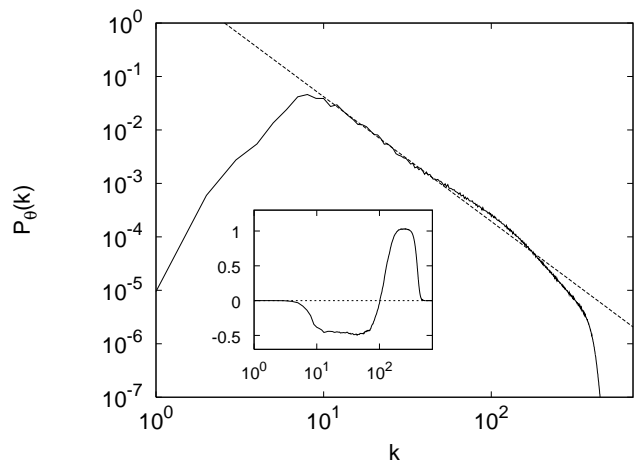}}
\caption{Power spectrum of the scalar field $\theta$ for the
asymptotic model $m \to 0$. Dashed line is the slope $k^{-7/3}$.
The inset shows the energy flux.} \label{fig:5}
\end{figure}

To study the limit $m \to 0$ let us write the l.h.s. of
(\ref{eq:1}) in $k$-space, and use the symmetry $j \leftrightarrow
k-j$:
\begin{eqnarray}& &
\frac{\partial \theta_{\bm k}}{m\partial t}  = \frac{1}{m}
\sum_{\bm j} [{\bm k},{\bm j}]j^{-m} \theta_{\bm j}\theta_{{\bm
k}-{\bm j}} \nonumber\\& &   = \frac{1}{2m} \sum_{\bm j} \left\{
[{\bm k},{\bm j}]j^{-m} + [{\bm k},{\bm k}-{\bm j}]|{\bm k}-{\bm
j}|^{-m} \right\}
\theta_{\bm j}\theta_{{\bm k}-{\bm j}} \nonumber\\
&& = \frac{1}{2m} \sum_{\bm j} [{\bm k},{\bm j}] \left\{ j^{-m} -
|{\bm k}-{\bm j}|^{-m} \right\} \theta_{\bm j}\theta_{{\bm k}-{\bm
j}} \label{eq:3}
\end{eqnarray}
where $[{\bm k},{\bm j}] = k_1j_2 -k_2j_1$, $k=|{\bm k}|$ and
$j=|{\bm j}|$. In the limit $m\to0$, equation~(\ref{eq:3}) has
still the form of a transport equation with the link between
$\theta$ and the stream function being $\psi({\bm k}) = - \ln|{\bm
k}| \theta({\bm k})$. Renormalizing $t \to m t$ one gets
\begin{equation}
\frac{\partial \theta_{\bm k}}{\partial t}
=
\frac{1}{2} \sum_{\bm j} [{\bm k},{\bm j}]
\ln (j/ |{\bm k}-{\bm j}|)
\theta_{\bm j}\theta_{{\bm k}-{\bm j}}+F+D
\label{eq:4}
\end{equation}
Numerical integration of eq.~(\ref{eq:4}) produces an inverse
turbulent cascade with power law spectrum $P_\theta(k) \sim
k^{-7/3}$. (see Figure~\ref{fig:5}.) In the range of scales of the
inverse cascade the field $\theta$ is self similar with scaling
exponent $h=2/3$, as confirmed by the re-scaling of the pdfs of
scalar increments (see Figure~\ref{fig:2}.).

%%%%%%%%%%%%%%%%%%%%%%%%%%%%%%%%%%%%%%%%%%%%%%%%%%%%%%

%\section{Statistical properties of zero-isolines}
%%%%%%%%%%%%%%%%%%%%%%%%%%%%%%%%%%%%%%%%%%%%%%%%%%%%%%%%%
\begin{figure} [t!]
\centerline{
\includegraphics[scale=0.7,draft=false]{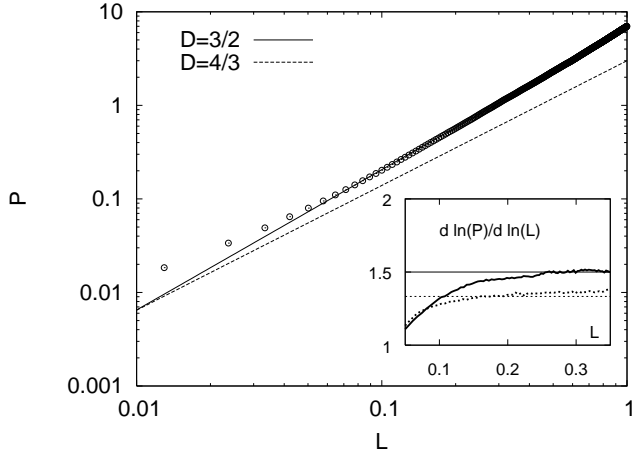}}
\caption{Perimeter $P$ of zero-isolines versus gyration radius $L$ for
$m=1/2$. Forcing length scale is $r_f \sim 0.06$.
In the inset we show the local slope of perimeter $P$ before
and after randomization of the phases of the scalar field
$\theta$ (solid and dashed line respectively).} \label{fig:6}
\end{figure}
%%%%%%%%%%%%%%%%%%%%%%%%%%%%%%%%%%%%%%%%%%%%%%%%%%%%%%%%%
\begin{figure} [b!]
\centerline{
\includegraphics[scale=0.7,draft=false]{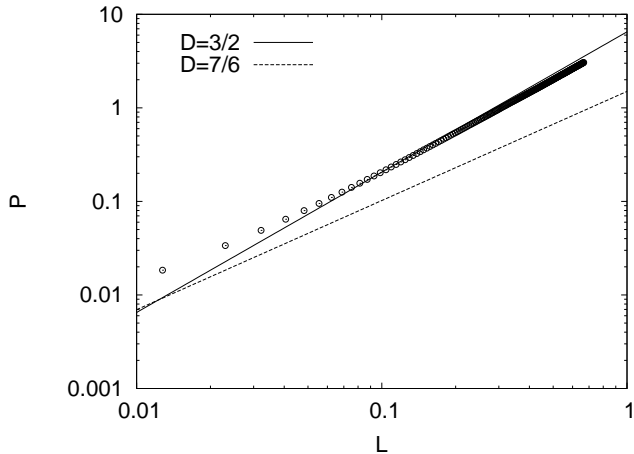}}
\caption{Perimeter $P$ of zero-isolines versus gyration radius $L$ for
$m \to 0$. Forcing length scale is $r_f \sim 0.06$.} \label{fig:7}
\end{figure}
%%%%%%%%%%%%%%%%%%%%%%%%%%%%%%%%%%%%%%%%%%%%%%%%%%%%%%%%%%%
%%%%%%%%%%%%%%%%%%%%%%%%%%%%%%%%%%%%%%%%%%%%%%%%%%%%%%%%%%%
\begin{figure} [t!]
\centerline{
\includegraphics[scale=0.7,draft=false]{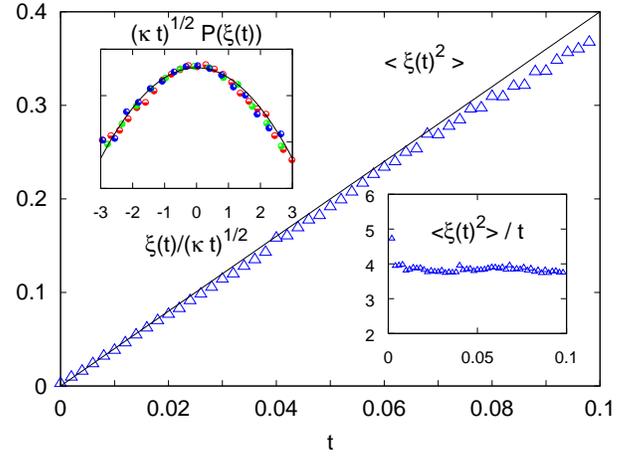}}
\caption{Statistics of the driving $\xi(t)$ for $m=1/2$}
\label{fig:8}
\end{figure}
%%%%%%%%%%%%%%%%%%%%%%%%%%%%%%%%%%%%%%%%%%%%%%%%%%%%%%%%%%%
\begin{figure} [b!]
\centerline{
\includegraphics[scale=0.7,draft=false]{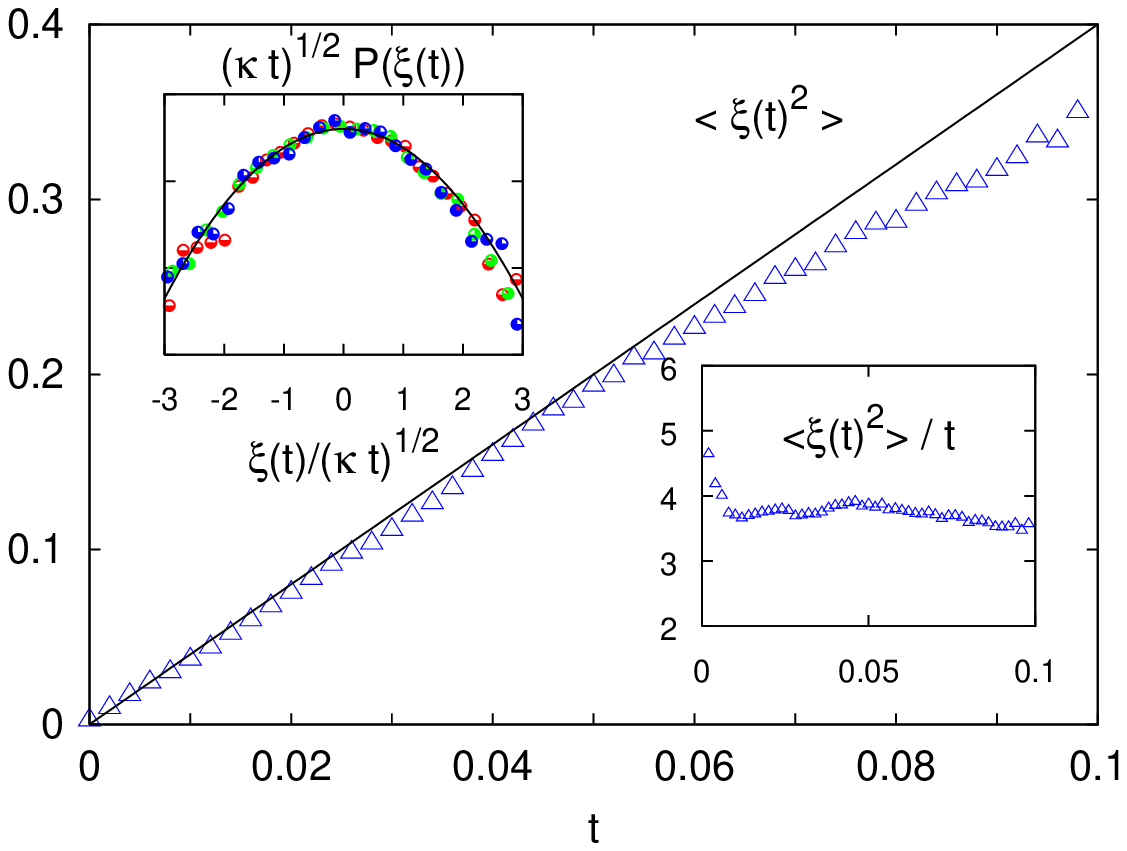}}
\caption{Statistics of the driving $\xi(t)$ for $m \to 0$}
\label{fig:9}
\end{figure}
%%%%%%%%%%%%%%%%%%%%%%%%%%%%%%%%%%%%%%%%%%%%%%%%%%%%%%%%%%%

Numerical investigation of Navier-Stokes (NS)
equation~\cite{BBCF06} and Surface Quasi Geostrophyc (SQG)
model~\cite{BBCF07} have shown that for two peculiar cases, namely
for $m=2,1$ the zero-isolines of the scalar field are
statistically equivalent to SLE i.e. could be mapped to 1d
Brownian walk. The class SLE is characterized by the respective
dimensionless diffusivity $\kappa$ \cite{Schramm,C05}. In
particular for NS the zero-vorticity isolines belong to the same
universality class of critical percolation and are equivalent to
SLE curves with $\kappa = 6$. For SQG the zero-temperature
isolines are SLE curves with $\kappa = 4$. It is therefore natural
to ask if the properties of conformal invariance for the
zero-isolines is a general property that holds for arbitrary
values of $m$.

To investigate this issue we consider the connected regions
of positive/negative sign of $\theta$, The boundaries of these clusters
are closed loops formed by the zero-$\theta$ isolines.

For $0< m \le 1$ the scalar $\theta$ is a self-similar rough field
with scaling exponent $0 < h < 1$. The relation between the
scaling exponent $h$ of a height function and the fractal
dimension $D$ of its isolines was suggested in \cite{KH95}:
$(3-h)/2=(7+2m)/6$ for $0<h<1$ and $0<m<1$.  Let us stress that
this is not the fractal dimension of the iso-set (known to be
equal to $2-h$ for $h>0$ and to $2$ for $h<0$) but that of a
single long isoline. One can thus conjecture the relation
$\kappa=4(1+2m)/3$. Indeed for $m=1$ it was found~\cite{BBCF07}
that the zero-isolines are SLE curves with $\kappa=4$, in
agreement with the above prediction. Nevertheless, in
Figures~\ref{fig:6} and~\ref{fig:7} we show the fractal dimension
of the isolines for $m=1/2$ and for the asymptotic model $m \to
0$. It both cases the fractal dimension measured is not agreement
with the prediction $D= (7+2m)/6$ but is compatible with $D=3/2$.

%%%%%%%%%%%%%%%%%%%%%%%%%%%%%%%%%%%%%%%%%%%%%%%%%%%%%%%%%%%
\begin{figure} [t!]
\centerline{
\includegraphics[scale=0.7,draft=false]{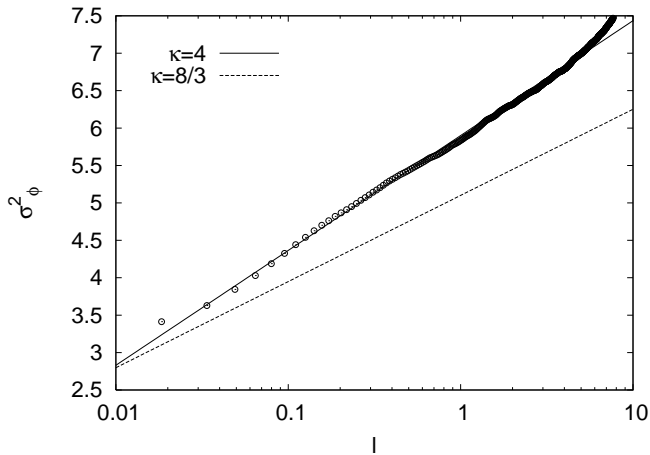}}
\caption{Variance of winding angle $\sigma^2_\phi$ as a function of the lenght $l$ along the isoline for $m=1/2$ } \label{fig:10}
\end{figure}
%%%%%%%%%%%%%%%%%%%%%%%%%%%%%%%%%%%%%%%%%%%%%%%%%%%%%%%%%%%
\begin{figure} [b!]
\centerline{
\includegraphics[scale=0.7,draft=false]{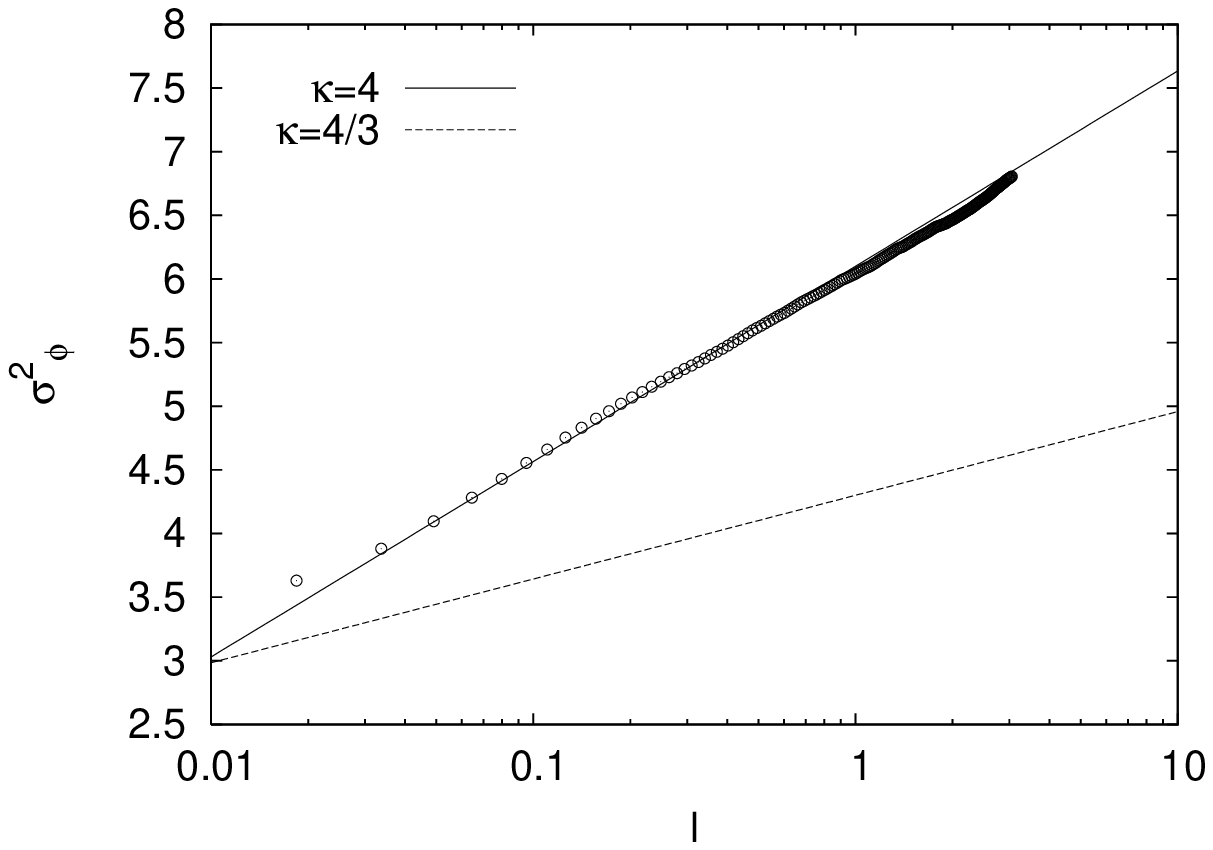}}
\caption{Variance of winding angle $\sigma^2_\phi$ as a function of the lenght $l$ along the isoline for $m \to 0$} \label{fig:11}
\end{figure}
%%%%%%%%%%%%%%%%%%%%%%%%%%%%%%%%%%%%%%%%%%%%%%%%%%%%%%%%%%%

Following the procedure described in~\cite{BBCF07} from the
zero-field lines we obtain an ensemble of curves in the the half
plane which are expected to converge in the scaling limit to
chordal SLE. Then we extract the driving $\xi(t)$ of the
corresponding Loewner equation. As shown in Figures~\ref{fig:8}
and~\ref{fig:9} the driving has Gaussian statistics with variance
$\langle\xi^2(t) \rangle \sim \kappa t$. For all the cases
considered with $0<m <1$ we found $\kappa=4$, which is in
agreement with the fractal dimension observed.

As a further test we study the statistics of the winding angle of
the zero-isolines. The winding angle $\phi$ is defined as the
degree with which the curve wind in the complex plane about a
point $w$ $\phi(z) = arg(z-w)$\cite{DB02, WW03}. The asymptotic
distribution for the winding angle at long distance $\ell$ along
the curve is Gaussian, with variance $\kappa /(4+\kappa/2)
\log(\ell)$. As shown in Figures~\ref{fig:10} and~\ref{fig:11} we
found that its variance grows like $2/3\log(\ell)$, thus
supporting the conjecture $\kappa=4$, and is not compatible with
the prediction $\kappa = (4+8m)/3$.

%%%%%%%%%%%%%%%%%%%%%%%%%%%%%%%%%%%%%%%%%%%%%%%%%%%%%%%%%%%%%%%%%%%%
\begin{figure} [t!]
\centerline{
\includegraphics[scale=0.7,draft=false]{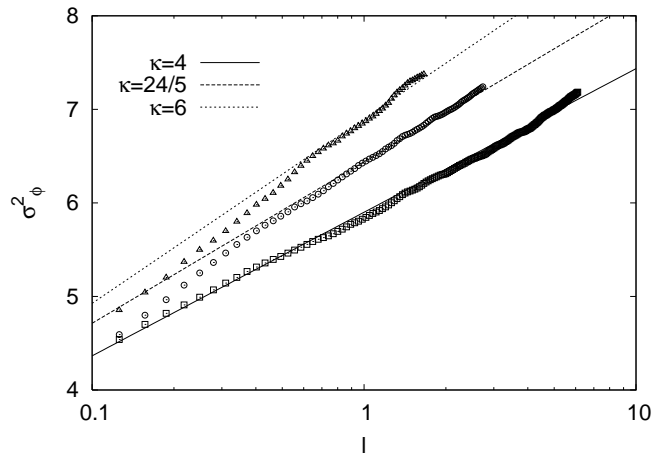}}
\caption{Variance of winding angle $\sigma^2_\phi$
as a function of the lenght $l$ along the isoline
for $m=1/2$ (squares),$m=3/2$ (circles), $m=2$ (triangles).}
\label{fig:12}
\end{figure}
%%%%%%%%%%%%%%%%%%%%%%%%%%%%%%%%%%%%%%%%%%%%%%%%%%%%%%%%%%%%%%%%%%%%%%%%

This model provides an example of non-trivial relation between the scaling
exponent of the field $\theta$ and the fractal dimension of its isolines. For
$0< m < 1$ the scaling exponent varies in the range $0 < h < 2/3$,  but the
fractal dimension remains constant $D=3/2$, at variance with what one would
expect from the relation $D=(3-h)/2$ which holds for Gaussian random field. The
crucial difference is that the scalar field $\theta$ is not random, but is the
result of turbulent dynamics.

Our findings can be understood in Lagrangian terms. The scaling exponent of the
velocity field is $h_v = (m-1)/3$. For $m > 1$ the scaling exponent is
positive, and therefore velocity difference scales as $\delta v(\ell) \propto
\ell^{(m-1)/3}$. Two Lagrangian trajectories moving in such velocity field
separate according to the Richardson law $\ell(t) \sim t^{3/(4-m)}$. Conversely
for $0< m < 1$, the  exponent is negative (that is the velocity is a
small-scale field like vorticity in Navier-Stokes), and velocity differences
are independent of the separation $\delta v (\ell) \simeq v_{rms}$. Lagrangian
trajectories will separate as $\ell(t) \propto t$. Perimeter $P$ and gyration
radius $L$ of clusters can be related by assuming that their ratio $P/L$, which
is proportional to the number of folds, grows as a random walk, i.e. as
$t^{1/2}$. Gyration radius grows as two-point distance $L(t) \propto \ell(t)$,
which gives $P \propto L t^{1/2} \propto L^{(10-m)/6}$ for $m \ge 1$ and $P
\propto L^{3/2}$ for $0< m <1$.

%%%%%%%%%%%%%%%%%%%%%%%%%%%%%%%%%%%%%%%%%%%%%%%%%%%%%%%%%%%%%%%%%%%%%%%%%%%%%%%
\begin{figure} [t!]
\centerline{
\includegraphics[scale=0.7,draft=false]{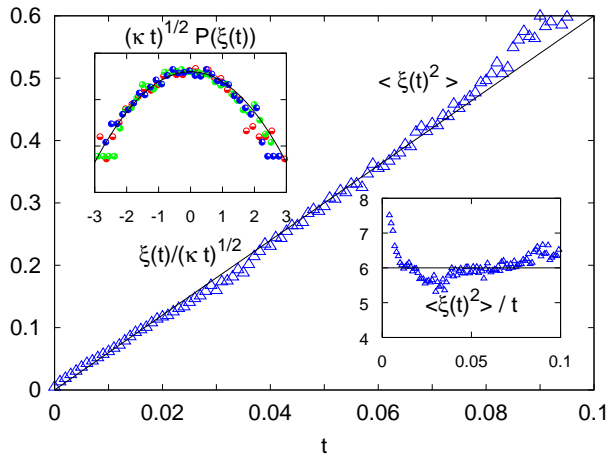}}
\caption{Statistics of the driving $\xi(t)$ for the stream-function isolines for $m=-2$}
\label{fig:13}
\end{figure}
%%%%%%%%%%%%%%%%%%%%%%%%%%%%%%%%%%%%%%%%%%%%%%%%%%%%%%%%%%%

The property of conformal invariance of the isolines is therefore
determined by the underlying dynamics of the field. As a test we
took the field $\theta$ and randomize its phases in Fourier space.
This procedure does not change the scaling exponent of the field,
but destroys all the correlations generated by the turbulent
dynamics. The isolines of this randomized field are no more
conformal  invariant, but their fractal dimension recover the
prediction for Gaussian random field (see inset of Figure~\ref{fig:6}).

Under the hypothesis that the isoline of the scalar field $\theta$ are SLE
curves one can obtain a conjecture for their universality class $\kappa$ from
the formula for the fractal dimension $D_* = 1+2/\kappa$ of the outer perimeter
$P$, which holds for $\kappa \geq 4$~\cite{D00}.
One obtains $\kappa =4$ for $0<m<1$ and $\kappa = 12/(4-m)$ for $m>1$,
which are in agreement with our findings ($m=3/2,1/2$ and $m \to 0$) and with
previous results ($m=1,2$).

Approach based on Schramm-Loewner Evolution provides a refreshingly
novel geometric insight into the statistics of turbulence and hints
at deep symmetry aspects of 2d flows which we are yet far from
understanding.

\end{document}